%% file: paper.tex
\documentclass[sigplan,10pt,nonacm]{acmart}

\input{macros}
\begin{document}

\title{Eliminate Branches by Melding IR Instructions}

\input{sections/abstract}

\author{Yuze Li}
\authornote{Equal contribution.}
\email{lyuze@vt.edu}
\affiliation{%
  \institution{Virginia Tech}
  \city{Blacksburg}
  \state{Virginia}
  \country{USA}
}

\author{Srinivasan Ramachandra Sharma}
\authornotemark[1]
\email{srinivasanr@vt.edu}
\affiliation{%
  \institution{Virginia Tech}
  \city{Blacksburg}
  \state{Virginia}
  \country{USA}
}
\author{Charitha Saumya}
\email{charitha.saumya.gusthinna.waduge@intel.com}
\affiliation{%
	\institution{Intel Corporation}
	\city{Santa Clara}
	\state{California}
	\country{USA}
}
\author{Ali R. Butt}
\email{butta@vt.edu}
\affiliation{%
  \institution{Virginia Tech}
  \city{Blacksburg}
  \state{Virginia}
  \country{USA}
}
\author{Kirshanthan Sundararajah}
\email{kirshanthans@vt.edu}
\affiliation{%
  \institution{Virginia Tech}
  \city{Blacksburg}
  \state{Virginia}
  \country{USA}
}
\maketitle




\input{sections/intro_compact}
\input{sections/background}

\input{sections/motivation}
\input{sections/overview}
\input{sections/design}
\input{sections/implementation}
\input{sections/evaluation}
\input{sections/discussion}
\input{sections/relatedwork}
\input{sections/conclusion}

\bibliographystyle{ACM-Reference-Format}
\bibliography{paper}

\end{document}

%% file: macros.tex
\usepackage{graphicx}
\usepackage{textcomp}
\usepackage[dvipsnames]{xcolor}
\colorlet{greenishgray}{gray!75!green}            
\usepackage{listings}
\usepackage{xspace}
\usepackage{amsmath,amsthm,verbatim,amscd}
\usepackage{caption}
\usepackage{subcaption}
\usepackage[ruled, vlined]{algorithm2e}

\usepackage{booktabs, multirow}
\usepackage{tikz}
\usepackage{pifont}
\usepackage{enumitem}
\usepackage{url}
\usepackage{wrapfig}
\usepackage[T1]{fontenc}
\usepackage{mdframed}
\usepackage{minted}
\usepackage{etoolbox}
\usepackage{multicol}

\definecolor{olivegreen}{rgb}{0.17, 0.53, 0.03}

\newtheorem{definition}{Definition}
\newcommand{\eg}{\textit{e.g.,}~}
\newcommand{\ie}{\textit{i.e.,}~}

\newcommand{\YL}[1]{{#1}}

\newcommand{\projFull}{\textsc{\textbf{M}\textbf{e}lding I\textbf{R} \textbf{I}ns\textbf{t}ructions}\xspace}
\newcommand{\proj}{\textsc{MERIT}\xspace}

\newcommand{\code}[1]{\texttt{#1}}

\def\BibTeX{{\rm B\kern-.05em{\sc i\kern-.025em b}\kern-.08em
    T\kern-.1667em\lower.7ex\hbox{E}\kern-.125emX}}
    
\newtoggle{bv}
\toggletrue{bv}

%% file: sections/abstract.tex
\begin{abstract}
Branch mispredictions cause catastrophic performance penalties in modern processors, leading to performance loss. While hardware predictors and profile-guided techniques exist, data-dependent branches with irregular patterns remain challenging. 
\YL{Traditional if-conversion eliminates branches via software predication but faces limitations on architectures like x86. It often fails on paths containing memory instructions or incurs excessive instruction overhead by fully speculating large branch bodies.}

This paper presents \projFull (\proj), a compiler transformation that eliminates branches by aligning and melding similar operations from divergent paths at the \textbf{IR instruction level}. By observing that divergent paths often perform structurally similar operations with different operands, \proj adapts sequence alignment to discover merging opportunities and employs safe operand-level guarding to ensure semantic correctness without hardware predication. Implemented as an LLVM pass and evaluated on 102 programs from four benchmark suites, \proj achieves a geometric mean speedup of 10.9\% with peak improvements of $32\times$ compared to hardware branch predictor, demonstrating the effectiveness with reduced static instruction overhead.
\end{abstract}

%% file: sections/intro_compact.tex
\section{Introduction}
\label{sec:intro}
Modern superscalar processors rely on speculative execution to maintain high instruction throughput. Conditional branches introduce a critical \textit{control-flow hazard}, breaking this linear flow and forcing the processor to guess the correct execution path. Mispredicting the branch paths can cause catastrophic performance penalties (up to 18\% IPC loss~\cite{branch_not_solved}). 

Hardware branch prediction has evolved significantly~\cite{tage, batage, tage_64, tage_sc_l, tage_sc_l_again}, and profile-guided approaches can eliminate specific problematic branches~\cite{whisper, thermometer}.
However, data-dependent branches with irregular patterns remain challenging due to the trade-off between prediction accuracy and silicon area.

A complementary software-based approach, \textit{if-conversion}, can eliminate the control-flow hazard altogether without any hardware changes~\cite{ifconversion, ifconversion_balance, predicated_execution, ir_ifconversion, wish_branches, early_ifconvert}, albeit the difficulty for compilers to identify which branches are destined to mispredict. 
Instead of generating a conditional jump, the compiler produces a single, straight-line sequence of predicated instructions. 
If-conversion operates at the {\bf Machine level IR} on a branch: when all instructions that are control dependent on a branch are predicated using the same condition as the branch, that branch can legally be removed.
\YL{However, this transformation suffers from two fundamental limitations on x86: First, x86 lacks the hardware predication support. If-conversion resorts to speculation-based approaches that execute \textit{both paths} and select between the final results. However, it cannot handle branches containing unsafe memory operations (loads from potentially invalid addresses or conditional stores), leaving many "convertible" branches untransformed.
Second, on x86, even when transformation succeeds, the overhead of duplicating all operations from both paths can outweigh the benefit of branch elimination, particularly for structurally-similar branches containing substantial computations.
}

This paper introduces \projFull (\proj), a compiler transformation that differs from traditional if-conversion by operating at the {\em target-independent IR level} rather than the target-dependent lower levels.
We observe that divergent control-flow paths often perform structurally similar operations differing primarily in their operands. 
Instead of speculatively executing entire paths, \proj aligns instruction sequences from both branches by identifying which operations are identical across paths.
These aligned instructions are melded into a single instruction with conditional operand selection, while unaligned instructions are melded after they are safely (\ie preserving correctness) matched with extraneous instructions inserted by \proj. 

This instruction-level approach achieves three key advantages over traditional if-conversion. 
(1) \textbf{Safe handling of memory operations through semantic analysis}: Unlike hardware predication or speculation-based if-conversion, \proj performs IR-level semantic analysis to reason about memory safety.
During melding, conditional loads and stores become unconditional loads and stores by getting correctly guarded by the branch condition to select the address and the value to guarantee correct results.
The availability of this semantic reasoning at the IR level enables \proj to transform branches that are traditionally skipped by if-conversion. 
(2) \textbf{Reduced static instruction overhead through IR instruction melding}: By merging structurally similar instructions rather than duplicating entire paths, \proj produces fewer operations. When both branches compute \code{x = a + offset} with different offsets, traditional if-conversion generates four operations (two additions, two selects) while \proj generates two (one select, one addition). This reduces code size, improves instruction cache utilization, and minimizes overhead that can negate branch elimination benefits.
(3) \textbf{Target-independent IR-level transformation}: 
The transformation is architecture-independent - the transformation is applicable irrespective of underlying architecture.
\proj produces branchless \code{select}-based code that unlocks downstream optimizations.
Straight-line code enables better instruction scheduling, register allocation, and vectorization—opportunities inhibited by control-flow boundaries. 
In order to see the performance benefit, this works focuses on x86 architecture.

\noindent Our key contributions are as follows:
\begin{itemize}[noitemsep,topsep=0pt,leftmargin=*]
    \item \proj: A novel compiler transformation that completely eliminates branches through instruction-level alignment and melding through semantic analysis, addressing the fundamental limitations of traditional if-conversion.
    \item Full Alignment Using Extraneous Instructions: \proj adapts the Smith-Waterman algorithm~\cite{smith_waterman} to discover locally optimal alignments across divergent paths, inserting safe extraneous instructions to fill unmergeable gaps and achieve full instruction sequence alignment.
    \item Correctness and Safety Guarantee: \proj ensures the data flow after inserting extraneous instructions will never corrupt the original program data flow. It uses semantic analysis to ensure safe memory access that traditional if-conversion cannot employ.
    \item We implement \proj as an LLVM pass and evaluate on 102 benchmarks from four benchmark suites, demonstrating a average speedup of 10.9\% with peak improvements of $32\times$, compared to pure hardware speculation.
\end{itemize}

In the rest of the paper we first present necessary background for this work (Section~\ref{sec:background}).
Then, we motivate our instruction-level approach with key examples in Section~\ref{sec:motivation} and illustrate the \proj transformation  design in Section~\ref{sec:design}.
We describe our implementation as an LLVM pass and its integration with profile-guided optimization (PGO) in Section~\ref{sec:implementation}, followed with a comprehensive evaluation in Section~\ref{sec:eval}.
Finally, we discuss limitations and future work in Section~\ref{sec:discussion}, mention related research in Section~\ref{sec:related}, and conclude in Section~\ref{sec:conclusion}.

%% file: sections/background.tex
\section{Background}
\label{sec:background}

\subsection{Control-flow Hazard and Branch Prediction}
Modern processors use deep pipelines and speculative execution to achieve high IPC. When encountering a conditional branch, the processor must predict its direction and speculatively fetch instructions from the predicted path. A correct prediction maintains pipeline flow, but a misprediction forces a complete pipeline flush and restart, a penalty that increases with pipeline depth.
Decades of researchers dived deep into branch prediction design to increase prediction accuracy. 
State-of-the-art hardware predictors are typically TAGE-like~\cite{tage, tage_64, tage_sc_l, tage_sc_l_again}, perceptron-based~\cite{perceptron_0, perceptron_1}, or a combination of those~\cite{perceptron_tage, tage_sc_l_again}.
Both types of predictors have distinct advantages: TAGE exploits the limited predictor storage very efficiently, whereas perceptron-based predictors can easily combine different sorts of input information. 
Recent researchers leverage PGO to increase prediction accuracy~\cite{whisper, thermometer, zangeneh2020branchnet}.
However, achieving high prediction accuracy often trades off with high space area on the chip. Plus, redesigning hardware to accommodate new compiler hints is non-scalable as it requires specialized ISA to accommodate software hints, hindering realistic in-production deployment.

\subsection{If-conversion}


If-conversion is a well-established compiler technique that eliminates branches by converting control-flow into data flow~\cite{ifconversion, early_ifconvert, ifconversion_balance, predicated_execution, ir_ifconversion,wish_branches}. 
The goal is to eliminate control dependencies that can limit the exposure of instruction-level parallelism (ILP) and avoid performance penalties from branch mispredictions.
Instead of a conditional jump, the compiler generates a single, straight-line sequence of predicated instructions.
The transformation is crucial for enabling other optimizations like vectorization and software pipelining, and it is especially critical for specialized architectures, such as DSPs.
Newest research applies the conversion early~\cite{early_ifconvert} to prevent the compiler from making optimizations that would harm performance.
On architectures like ARM, this uses predicates to conditionally commit instruction results. 
The hardware executes these instructions, but only allows the results to be committed if their governing predicate is true; otherwise, the instruction is nullified.

\YL{However, x86 lacks support for hardware predication, it does speculative execution, which relies on safe execution on both paths. Consequently, compiler does not apply if-conversion on branch paths containing memory instructions, leaving numerous potential optimizable code on the table.}
\YL{Secondly, if-conversion transforms naively: it must fully execute both paths before selection.}
This "execute everything, select results after" model can introduce excessive instruction overhead, especially for branches with large bodies, often degrading performance despite eliminating mispredictions.
To mitigate this, various approaches have been proposed, such as using PGO to selectively apply the transformation~\cite{pgo_ifconv, pgo_predication}. 
Other hybrid techniques, like Wish Branches~\cite{wish_branches}, defers the decision to runtime by encoding both the branch and its if-converted version in the binary.
Despite these mitigations, the inefficient full-speculation model is largely retained. 
In contrast, \proj moves beyond this branch-level decision by operating at the IR instruction level, melding similar operations to reduce redundant work rather than fully speculating both paths.

\subsection{Sequence Alignment}
\proj adapts the Smith-Waterman algorithm~\cite{smith_waterman}, originally developed for biological sequence alignment, to align structurally similar instruction sequences across divergent control-flow paths. 
The algorithm uses dynamic programming to find optimal local alignments by scoring matches 
(similar instructions), mismatches (dissimilar instructions), and gaps (instructions present in only one path), enabling \proj to discover merging opportunities even when instructions appear in different orders or with intervening operations (detail in Sec.~\ref{sec:design}).

%% file: sections/motivation.tex
\section{Motivation}
\label{sec:motivation}

In this section, we use two examples to demonstrate how \proj eliminates branches by melding (and adding extraneous) instructions on both paths.

\subsection{How \proj Eliminates Branches}
The primary intuition behind \proj transformation is to insert extraneous (\ie dummy) instructions to make the same operations on divergent paths and to avoid the explicit branch.
Consider the function \code{to\_upper} in Listing~\ref{lst:to_upper}
\footnote{\YL{Only for illustration purpose. Standard compiler optimizations happen \textbf{after} \proj's transformation.}}, which converts all lowercase letters in a given string of length \code{SIZE} to uppercase ones. 
The \code{if}-conditional within the \code{for}-loop is executed repeatedly while iterating through each character of the string. This branch is highly unpredictable because of the high data dependency on the string. 
Experimenting \code{to\_upper} on x86 shows that around 27\% of all dynamic branches in the function are mispredicted and it leads to an extremely low IPC of 0.48.

\begin{listing}[ht]
\begin{lstlisting}[numbersep=1mm, language=C, basicstyle=\ttfamily\small, numberstyle=\ttfamily\footnotesize, stringstyle=\color{purple}, keywordstyle=\color{OliveGreen}\bfseries]
void to_upper(char *str) {
    for(int i=0; i<SIZE; i++) {
        if(str[i]>='a' & str[i]<='z')
            str[i] = str[i] + 'A' -'a';
    }
}

void to_upper_branchless(char *str){
    for(int i=0; i<SIZE; i++){
        bool cond = (str[i]>='a') & (str[i]<='z');
        unsigned int diff = cond ? 'A'-'a' : 0;
        str[i] += diff;
    }
}
\end{lstlisting}
\caption{Motivating Example}
\label{lst:to_upper}
\end{listing}

The function \code{to\_upper\_branchless} is semantically the same as \code{to\_upper}.
However, the \code{if}-conditional within the loop is eliminated and the loop body is straight line code without any branches (here the ternary operator would be translated to \code{cmov} instructions on x86 assembly).
This version has \textit{nearly zero} branch mispredictions and yields a $32\times$ speedup compared to the original version.
In this version, irrespective of the value \code{cond}, which holds a value representing the outcome of the branch in the original version, the computation of adding a constant value \code{diff} to the character of the string is executed.
This does not affect the correctness, as when the branch is not taken, the transformation only adds a zero to the character.
\proj automatically recognizes the \code{branched} code and transforms it into the \code{branchless} version. 
\YL{Note that it appears \code{str[i] += diff} executes unconditionally in every iteration. Segmentation fault might occur if \code{diff} is \code{0}, when \code{str[i]} is read-only. However, as we will later show in Figure~\ref{fig:codegen_step3}, \proj internally writes to a safe, dummy memory location if fallback path is taken, rather than writing to the potential faulting address.}


\subsection{How \proj Differs from if-conversion}
\begin{figure}[t]
    \centering
    \includegraphics[width=0.9\columnwidth]{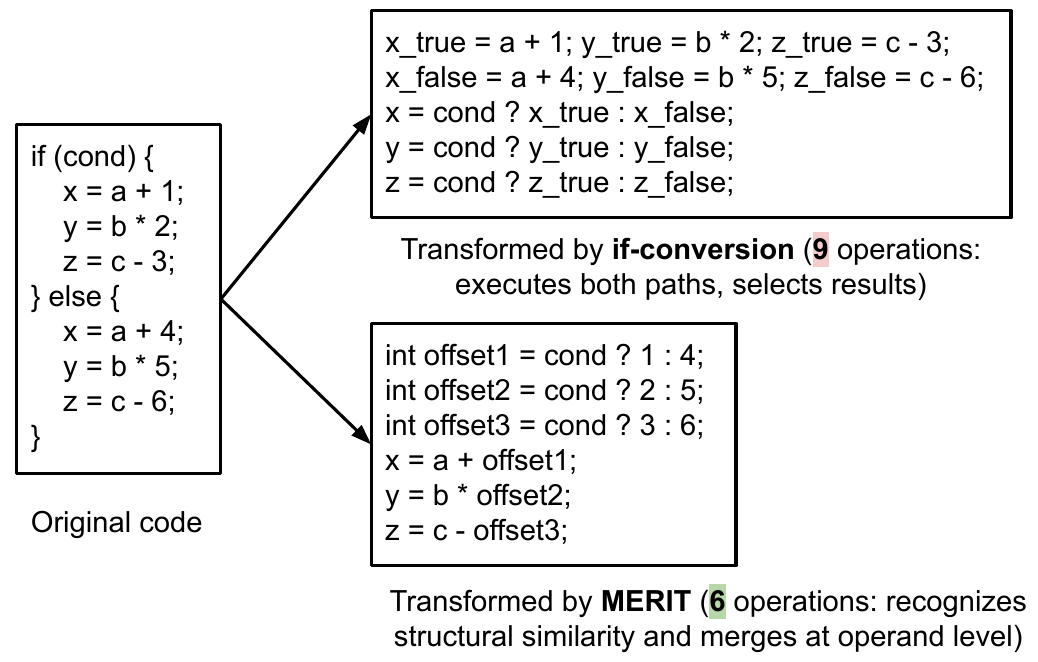}
    \caption{The given example shows \proj can save 33\% operations \YL{compared to if-conversion} by recognizing that both paths perform the same operations with different constants, enabling instruction-level merging rather than result-level selection.}
    \label{fig:motivation}
\end{figure}

Figure~\ref{fig:motivation} shows a case where both branches perform structurally identical computations differing only in their operands.
\YL{On commodity architectures that rely on \code{cmov} (x86),} if-conversion uses a speculative "compute everything, select results" method, calculates the outcomes for both the if and else paths entirely, and then uses (\code{cmov}) to select the correct final results.
While this strategy effectively eliminates branch misprediction penalties, it introduces redundant work by performing every arithmetic operation twice when only one outcome is needed.

On the contrary, \proj uses instruction-level alignment to recognize the structural similarity between the branches. Instead of duplicating the computations, it performs operand-level merging. This "select operands, compute once" method first uses \code{select} instructions to choose the correct constant based on the branch condition, and then executes each arithmetic operation a single time with the selected operand. \proj achieves a significant reduction in total operations by \textit{merging the computation at the instruction level, rather than merely selecting between fully-computed results}, leading to better performance.

%% file: sections/design.tex
\section{Detailed Design}
\label{sec:design}
In this section, we describe the compiler transformation \proj to statically merge divergent control-flow paths of a program.
We illustrate the phase of inserting {\em extraneous instructions} to make the sequence of operations in the instructions of the \code{if-then-else} statements identical.
The insertion of extraneous instructions helps to fully meld the sequences of instructions, resulting in the elimination of expensive branches (\ie hard to predict branches).
We also provide a proof sketch that \proj is a semantic-preserving transformation.

\subsection{\proj Transformation}
\label{sec:transformation}

The \proj transformation is based on similar principles as DARM~\cite{DARM}, which is a compiler optimization that improves the performance of GPU programs by statically merging control divergent program paths.
DARM mainly identifies \code{if-then-else} branches with similar basic blocks (\ie isomorphic control flow regions) and merges the common instructions within those basic blocks into convergent ones.
The main focus of DARM is not completely eliminating branches; it happens to eliminate branches if the operations in the instruction sequences are identical.
However, it is a rare occurrence to get instruction sequences with identical operations on both sides of a branch.
In case of non-identical portions, DARM puts them in separate basic blocks to conditionally execute them, and this will increase the number of branches.

The key idea of \proj is inserting extraneous instructions into both sides of a conditional branch such that the sequences of operations become identical. 
An extraneous instruction is needed when the instruction alignment contains an {\em unaligned instruction}. 
We can formally define the alignment of instruction sequences by using the following definitions.
Consider a program with an \code{if-then-else} branch with two basic blocks $B_t$ and $B_f$ (\ie diamond-shaped control-flow), and $I_t$ and $I_f$ are the instruction sequences of those basic blocks, respectively. 

\begin{definition}
	\textbf{Compatible Pair:} Let $(a, b)$ be a pair of instructions. It is a compatible pair if and only if the operation of $a$ and $b$ is identical (\eg operation of $a$ and $b$ are \code{iadd}). It is not necessary for instructions $a$ and $b$ to have identical operands for the pair $(a, b)$ to be compatible.
\end{definition}

\begin{definition}

  \textbf{Instruction Alignment:} \YL{Let $I_t = \{i_1^t < \dots < i_n^t\}$ and $I_f = \{i_1^f < \dots < i_m^f\}$ be the \textbf{complete} ordered sequences of instructions in basic blocks $B_t$ and $B_f$ respectively. An instruction alignment is an ordered sequence of pairs $A = \{(a_1, b_1) < \dots < (a_k, b_k)\}$ such that $a_j \in I_t \cup \{\psi\}$, $b_j \in I_f \cup \{\psi\}$. The sequence length $k$ satisfies $\max(n, m) \le k \le n + m$, and $\forall j \in [1, k]$, $(a_j, b_j) \neq (\psi, \psi)$. If $a_j \neq \psi$ and $b_j \neq \psi$, then $(a_j, b_j)$ is a compatible pair for merging. Here, $\psi$ denotes the absence of an instruction (i.e., empty slot).}
\end{definition}

\begin{definition}
  \textbf{Unaligned Instruction:} Let $(a_i, b_i) \in A$ be a pair in an instruction alignment $A$ such that $a_i = \psi$ or $b_i = \psi$. 
  Let $i'$ be the valid instruction in the pair $(a_i, b_i)$, then $i'$ is called an unaligned instruction.
\end{definition}

\begin{definition}
  \textbf{Complete Alignment:} 
  \YL{An instruction alignment $A'$ is called \textbf{complete} if it does not contain any unaligned instructions (i.e., for all pairs $(a_j, b_j) \in A'$, $a_j \neq \psi$ and $b_j \neq \psi$). This implies that the lengths of the instruction sequences in the aligned basic blocks must be equal ($n = m$).}
\end{definition}

If the instruction alignment for $B_t$ and $B_f$ is {\em complete}, we can fully merge $B_t$ and $B_f$ into a single basic block, eliminating the conditional branch.
The first step of \proj is to transform the alignment $A$ into a complete alignment $A'$.

\begin{definition}
  \textbf{Extraneous Instruction:} 
  \YL{Let $I_t$ contain an unaligned instruction $i'$. An instruction $i''$ is inserted into $I_f$ such that $i'$ and $i''$ are compatible and form a pair in the complete alignment $A'$. $i''$ is defined as an \textbf{extraneous instruction}. While functionally similar to speculative predication, MERIT generates $i''$ using \textbf{semantic safety analysis} to ensure it does not trigger exceptions (e.g., division-by-zero, memory traps) when executed on the path where it was originally absent.}
\end{definition}

Extraneous instructions are essential for a complete alignment and elimination of the branch.
Nevertheless, it is important to ensure that the transformation does not change the semantics of the program.

\paragraph{Select minimization.}
In the code generation process of \proj, extra \code{select} operations are inserted if the operands of the two merged instructions do not match.
This process can increase the number of instructions in the program, and extra \code{select} operations can make the data-flow more complex.
Therefore, minimizing the number of \code{select} instructions generated by \proj transformation is essential.
\code{Select} operations can be minimized if both sides of the conditional branch have similar def-use chains.
More precisely, let $i_t = op(o_t^1, o_t^2)$ and $i_f = op(o_f^1, o_f^2)$ be two aligned instructions in the alignment $A$. 
Merging $i_t$ and $i_f$ does not require additional \code{select} operations if $o_t^1 = o_f^1$ and $o_t^2 = o_f^2$ or $(o_t^1, o_f^1)$ and $(o_t^2, o_f^2)$ are also produced by aligned instructions in the $A$. 
\YL{For example, consider two aligned instructions $i_t$: \code{r1 = t1 + 4} and $i_f$: \code{r2 = t2 + 4}. If $t1$ and $t2$ are the results of a previously aligned and merged instruction pair (producing a single value $t_{new}$), and the constant operand 4 is identical, \proj generates a single merged instruction \code{r\_new = t\_new + 4}. No additional select instructions are required because the operands are either shared constants or flow directly from the already-merged data dependency chain.}

\paragraph{\YL{IR alignment scoring.}}
\YL{To quantify the similarity of def-use chain from both paths, \proj introduces the following formula: $Score = num\_matches \times match\_bonus - num\_gaps \times gap\_penalty$. Essentially, the score represents the net benefit of matching instructions (without introducing extra \code{select} operation).
If the score falls below the default threshold (0.2, can be tuned during compilation), \proj rejects the transformation. }

\paragraph{Operands for extraneous instruction.}
There is some flexibility in setting operands for extraneous ALU instructions.
We can set all the operands of the extra instruction to some safe, constant value depending on the semantics of the instruction (\eg $0$ for \code{add} and $1$ for \code{div} instructions).
On the other hand, we can set operands such that def-use chains are preserved and \code{selects} are minimized.
In \proj, we perform a mix of both.
We preserve the def-use chains and minimize \code{select} operations if the extra instructions cannot result in failures (such as overflow, underflow, division by zero, or undefined behavior). 
The operand setting process is explained with an example at the end of this section.

We follow the following criteria when aligning memory operations.
\begin{itemize}[noitemsep,topsep=0pt,leftmargin=*]
  \item If we can determine two aligned memory operations accessing the same address at compile time, we merge them into a single memory operation (no select needed). 
  \item If two aligned memory operations access different memory locations, we merge them and select the address for the merged instruction conditionally.
  \item If there is an unaligned memory operation, we will insert an extraneous memory instruction to the aligned empty slot to access a location from the properly initialized {\em safe global memory space} defined by \proj. This safe global memory space is allocated and initialized by the compiler and guaranteed not to alias to any other memory location in the program. 
  \item \YL{For thread safety, the compiler prepares separate memory space for each thread.}
\end{itemize}

\begin{figure*}[t]
  \begin{mdframed}
  \begin{subfigure}[b]{0.24\textwidth}
  \begin{lstlisting}[numbersep=1mm, language=C, basicstyle=\ttfamily\footnotesize, numberstyle=\ttfamily\footnotesize, stringstyle=\color{purple}, keywordstyle=\color{OliveGreen}\bfseries]
  // ...
  if ((text[i] >= 'a') 
    & (text[i] <= 'z')) {
    t1 = text[i] - 'a';
    t2 = t1 + 'A';
    text[i] = t2;
  } else {
  }
  \end{lstlisting}
  \caption{\YL{\code{to\_upper} function with an empty \code{else} section inserted.}}
  \label{fig:codegen_step1}
  \end{subfigure}
  \hfill
  \begin{subfigure}[b]{0.24\textwidth}
  \begin{lstlisting}[numbersep=1mm, language=C, basicstyle=\ttfamily\footnotesize, numberstyle=\ttfamily\footnotesize, stringstyle=\color{purple}, keywordstyle=\color{OliveGreen}\bfseries]
  // ...
  if ((text[i] >= 'a') 
    & (text[i] <= 'z')) {
    t1 = text[i];
    t2 = t1 - 'a';
    t3 = t2 + 'A';
    text[i] = t3;
  } else {
    t5 = *mem;
    t6 = t5 - 0;
    t7 = t6 + 0;
    *mem = t7;
  }
  \end{lstlisting}
  \caption{\YL{Code after extraneous code insertion.}}
  \label{fig:codegen_step2}
  \end{subfigure}
  \hfill
  \begin{subfigure}[b]{0.4\textwidth}
  \begin{lstlisting}[numbersep=1mm, language=C, basicstyle=\ttfamily\footnotesize, numberstyle=\ttfamily\footnotesize, stringstyle=\color{purple}, keywordstyle=\color{OliveGreen}\bfseries]
  // ...
  unsigned is_lower = 
    (text[i] >= 'a') & (text[i] <= 'z');
  t1_t5  = is_lower == 0 ? text[i] : *mem;
  s1     = is_lower == 0 ? 0 : 'a'; 
  t2_t6  = t1_t5 - s1;
  s2     = is_lower == 0 ? 0 : 'A'; 
  t3_t7  = t2_t6 + s2;
  t4_t8  = is_lower == 0 ? &text[i] : mem; 
  *t4_t8 = t3_t7;
  \end{lstlisting}
  \caption{\YL{Code after the merging step.}}
  \label{fig:codegen_step3}
  \end{subfigure}
  \end{mdframed}  
\caption{\proj transformation example}
\label{fig:codegen}
\end{figure*}

\paragraph{Example.}
Now we explain how \proj transformation works in action using our running example (Listing~\ref{lst:to_upper}).
Figure~\ref{fig:codegen} shows how \code{to\_upper} function is transformed at each stage.
Figure~\ref{fig:codegen_step1} shows \code{to\_upper} function with an empty \code{else} section inserted. 
This is an extra canonicalization step of \proj that converts \code{if-then} to \code{if-then-else} form, which allows it to merge \code{if-then} branches.
Also, instructions are shown on separate lines (Lines 4-6) for better readability.
Figure~\ref{fig:codegen_step2} shows the code after extraneous code insertion. 
Here, the \code{else} path is empty; therefore, all the instructions are unaligned.
The \code{else} path contains inserted extraneous instructions.
For example, the load operation in Line 4 is repeated with a location (\code{mem}) from {\em safe global memory space} in Line 9 after the extraneous code insertion.
\proj also tries to preserve def-use chains and minimize select operations required for merging.
For example, variable \code{t7} at Line 11 uses \code{t6} at Line 10.
This is similar to variable \code{t3} using \code{t2} as its first operand.
\code{t7} uses \code{0} as its second operand to avoid any overflow/underflow bugs.
This example also demonstrates how store instructions are handled during extraneous code insertion.
On \code{if} path, there is a store (Line 7) of value \code{t3} to \code{text[i]}.
On \code{else} path, the same store is performed (Line 12), but the stored location is \code{mem} and the stored value is \code{t7}.
Figure~\ref{fig:codegen_step3} shows the code after the merging step
\footnote{\YL{Verbose sequence to illustrate step-by-step melding process. Standard passes (Constant propagation, DSE) would run after \proj to clean up constants.}}.
Notice extra select instructions (shown as a ternary operator) are inserted to select operands if input operands do not match.
The transformed program is much faster to execute than the original one (Section~\ref{sec:motivation}).

\paragraph{Putting all together.}
We describe the overall \proj transformation for a program in Algorithm~\ref{alg:cfm}.
\proj transformation iterates through all the functions in a program.
For each function, it collects all the valid branches for applying the \proj transformation.
The {\em structural validity} of a branch is determined by the two paths of the branch having straight-line control-flow converging at a basic block. 
In other words, control-flow regions of \code{if-then-else} (\ie two-sided branches) or \code{if-then} (\ie one-sided branches).
Then, for each valid branch, it computes the instruction alignment for the two sides of the branch.
If the alignment cannot be made complete, it inserts extraneous instructions to complete the alignment.
In the degenerate case of one-sided branches (\code{if-then}), a whole basic block of extraneous instructions will be added.
Finally, it merges the two blocks and simplifies the function, removing the branch.
We repeatedly perform these steps until there are no changes to the function.
Not all structurally valid branches require \proj transformation.
The decision of which valid branches need the transform is flexible and best guided as a Profile-guided Optimization (PGO) to selectively target challenging and hard-to-predict branches (Sec.~\ref{sec:pgo}).

\begin{algorithm}[t]
  \SetAlgoLined
  \SetKw{Continue}{continue}
  \SetKw{False}{false}
  \SetKw{True}{true}
  \SetKw{Is}{is}
  \KwIn{Original Program $P$ and Excluded Locations $L$}
  \KwOut{Transformed Program $P'$}
  \caption{\proj Transformation Algorithm}
  \label{alg:cfm}

   \For{F $\in$ functions(P)}{
   	Changed $\gets$ \False\;
   	\Repeat{Changed \Is \False}{
   		ValidBranches $\gets$ collectValidBranches(F, L)\;
	   	\For{BI $\in$ ValidBranches}{
	   		BL, BR $\gets$ getDiamondBlocks(BI)\;
	   		A $\gets$ computeAlignment(BL, BR)\;
	   		\If{canCompleteAlignment(A, BL, BR) \Is \False}{
	   			\Continue
	   		}
	   		BL', BR' $\gets$ insertExtraInsts(A, BL, BR)\;
	   		mergeBlocks(BL', BR')\;
	   		Changed $\gets$ \True\;
	   		
	   		\If{Changed \Is \True}{
	   			simplify(F)\;
	   		}
	   	}
   	}
   }
\end{algorithm}

\paragraph{Correctness of \proj}
The \proj transformation is a semantic-preserving transformation, and it does not introduce any crashes to the program. 
Without loss of generality, let us reason that applying \proj to a single branch does not affect the correctness.
If the sequences of instructions on both sides of the branch are identical in terms of operation, melding those sequences with \code{select} instructions is semantic preserving.
It is enough to reason that the insertion of {\em extraneous instruction} does not affect the correctness of the program.
We can prove that any extraneous instruction does not affect the original data flow of the program since the values produced by extraneous instructions would never be consumed by the original instructions of the program.
Also, extraneous instructions should never write to a memory location of the program.
Hence, the data flow of the program stays intact even after the insertion of extraneous instructions. 
In addition, an extraneous instruction does not access a memory location that is not supposed to be accessed in the path where the instruction is inserted. 
Therefore, \proj transformation does not affect the correctness of the program or introduce crashes.

%% file: sections/implementation.tex
\section{Implementation}
\label{sec:implementation}




\subsection{Compiler Integration}
{We implement \proj as an LLVM pass in LLVM-14, integrated into the standard pipeline and enabled by default at optimization levels above \code{-O0}. It is strategically scheduled after early canonicalization and simplification passes (\eg SimplifyCFG, SROA, and Mem2Reg). This position is intentional: it allows \proj to operate atop LLVM's existing CFM optimizations. Furthermore, early simplification passes like Mem2Reg remove redundant code, making the IR smaller and simpler. This reduces the compile-time complexity for \proj's instruction alignment step. By running before backend optimizations, \proj also avoids potential interference with later-stage alias analysis or memory layout changes.}

\subsection{Enabling Profile-guided Optimization}
\label{sec:pgo}
While \proj can achieve high performance by instruction-level merging on control-flow patterns, indiscriminately applying the transformation across all branches in a program can lead to poor performance. 
The overhead introduced by \code{select} instructions and operand multiplexing may exceed the benefit of branch elimination in cases where branch prediction accuracy is already high or when the merged instruction sequence is significantly longer than the original branching code. 
We address this challenge by enhancing the application of \proj in the fashion of a Profile-guided Optimization (PGO): selectively transform based on observed empirical performance.

\proj provides fine-grained filtering mechanisms accessible through compiler command-line options, either filtering by source file names, function names, or line numbers to enable selective application of the transformation. 
For instance, developers can specify function-level inclusion or exclusion lists using \url{-include-func-names=func1,func2} to selectively enable \proj only for high-value targets (\eg hot functions or functions with highest number of unpredictable), or \url{-exclude-func-names=func3,func4} to blacklist problematic functions (\eg functions known to cause heavy overhead).
Additionally, \proj supports file-level filtering via \url{-exclude-file-names=utils.c,legacy.c} to exclude entire source files, and line-level precision through \url{-json-include-lines=profile.json} where the JSON file maps source filenames to arrays of line numbers that should undergo transformation.
This multi-granularity filtering infrastructure enables developers to iteratively refine their optimization strategy by excluding transformation sites that exhibit poor performance characteristics while retaining those that demonstrate clear benefits.


%% file: sections/evaluation.tex
\section{Evaluation}
\label{sec:eval}


\subsection{Experimentation Setup}
\paragraph{Hardware and OS}
All tests are run on a x86 server with Intel Xeon(R) Silver 4314 CPU, featuring 32 cores, and 192 GB of RAM. \YL{The microarchitecture is Intel Sunny Cove (Ice Lake) with TAGE-based branch predictor.} The system runs Ubuntu 22.04.5 LTS based on kernel version 6.8.0.

\paragraph{Workloads} 
We evaluate \proj on totally 102 benchmarks from four benchmark suites: 
\begin{itemize}[noitemsep,topsep=0pt,leftmargin=*]
    \item 24 branch-heavy \code{microbenchmarks}, ranging multiple common algorithms.
    \item 16 CPU SPEC2017~\cite{cpu2017} benchmarks compiled in \code{rate} mode. \YL{We intentionally exclude all Fortran workloads because LLVM-14 lacks the \code{flang-new} code generation capabilities required to lower Fortran to LLVM IR.}
    \item 40 \code{pyperformance}~\cite{pyperformance} benchmarks invoked from \proj-optimized Python interpreter.
    \item 22 \code{TPC-H}~\cite{tpch} OLAP queries processed by \proj-optimized SQLite~\cite{sqlite} engine.
\end{itemize}

\paragraph{Comparison targets} 
We compile the benchmarks into four modes for evaluation:
\begin{itemize}[noitemsep,topsep=0pt,leftmargin=*]
    \item \textbf{\proj}: The transformation is applied, with backend optimization disabled.
    \item \textbf{\proj-PGO}: PGO is applied to exclude those functions exhibiting performance degradation(Sec.~\ref{sec:pgo}) based on heuristic results.
    \item \textbf{\proj-O2}: The transformation is applied, with backend optimization enabled.
    \item \textbf{if-conv}: Early if-conversion~\cite{early_ifconvert} is applied, with backend optimization enabled.
\end{itemize}
The reported performance for all modes is normalized to its corresponding baseline (pure hardware prediction without any special compiler transformation).
To ensure a fair comparison, this baseline is always compiled with the same backend optimization level as the specific mode being evaluated. Specifically, \proj and \proj-PGO are compared to a baseline with backend optimizations disabled, while \proj-O2 and if-conv are compared to a baseline with backend optimizations enabled. \YL{We chose the median as each reported number across 5 independent runs, with negligible variances.}

\subsection{Microbenchmarks}
\begin{figure}[t]
    \centering
    \begin{subfigure}{\columnwidth}
        \centering
        \includegraphics[width=0.9\columnwidth]{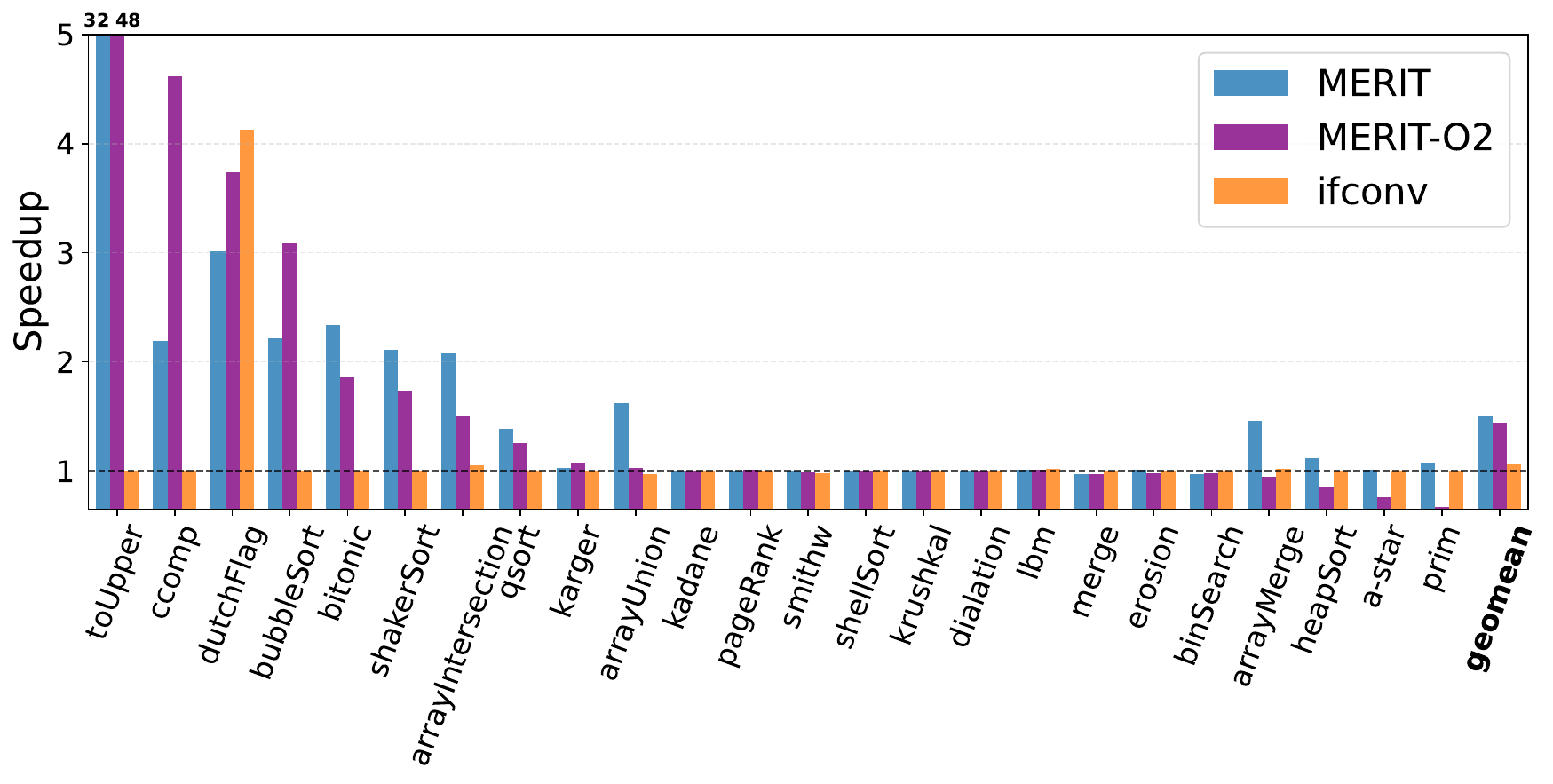}
        \caption{\proj achieves performance geomean of 1.51 compared to if-conversion's 1.06.}
        \label{fig:dse_perf}
    \end{subfigure}
    \vspace{1em}
    \begin{subfigure}{\columnwidth}
        \centering
        \includegraphics[width=0.9\columnwidth]{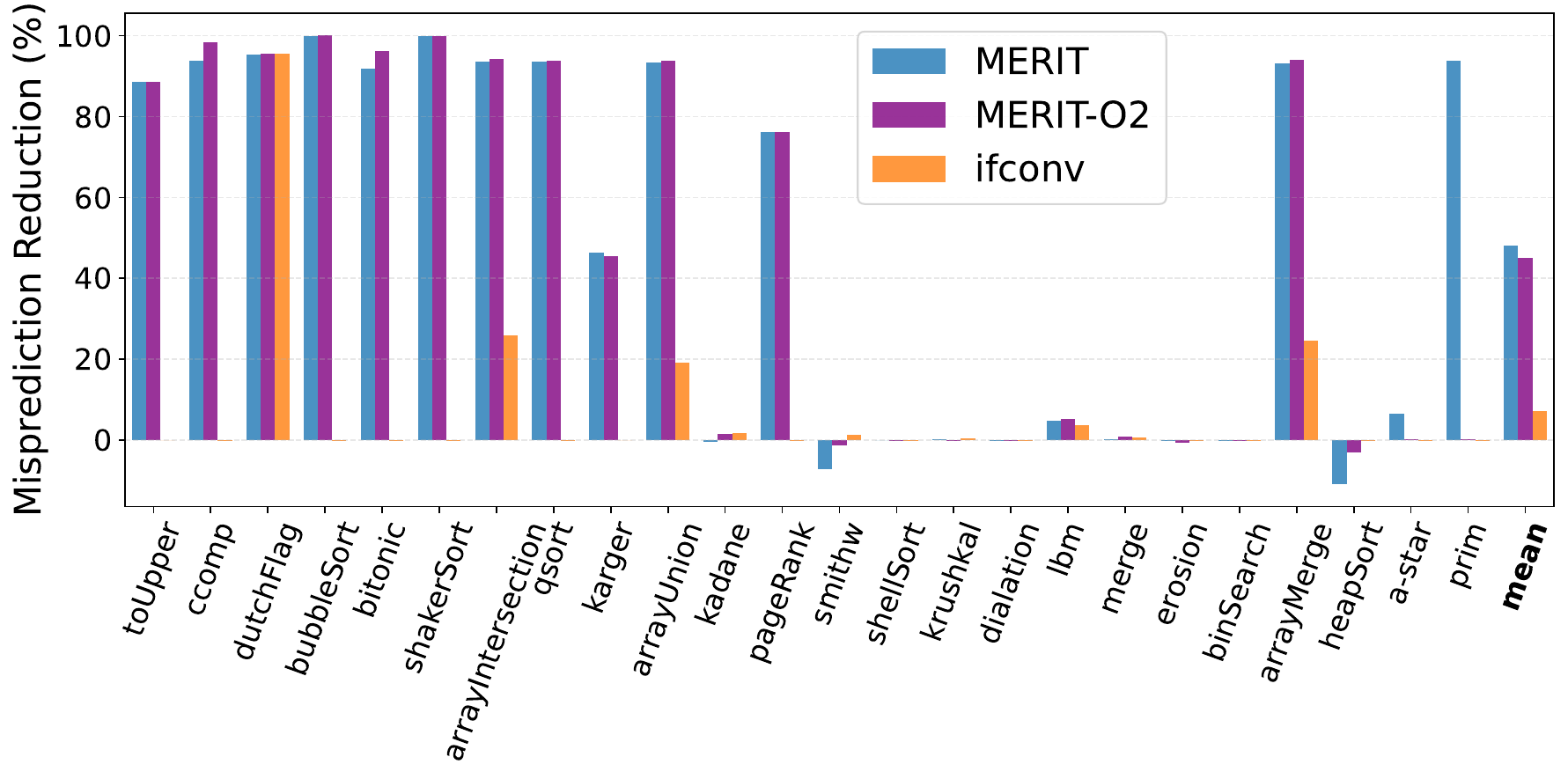}
        \caption{\proj achieves average 48\% compared to if-conversion's 7.2\% branch misses reduction.}
        \label{fig:dse_br_miss_reduction}
    \end{subfigure}
    \vspace{-2em}
    \caption{Performance and branch misses reduction of \proj and early if-conversion on the \textbf{microbenchmarks} compared to hardware speculation (no transformation).}
    \label{fig:dse_perf_br_miss}
\end{figure}

\paragraph{Backend optimization impact on microbenchmarks} 
In Figure~\ref{fig:dse_perf}, we measure the run time performance of 24 branch-heavy microbenchmarks ranging from sorting, comparison, traversing, etc. 
\proj can achieve up to $32\times$ speedup than hardware speculation in \code{toUpper}, with an 1.51x geometric mean (geomean) speedup across all benchmarks.
\proj-O2 achieves better performance than \proj in \code{toUpper}, \code{ccomp}, \code{dutchFlag}, and \code{bubbleSort} but fails in others (geomean 1.44x).
Notice that in \code{arrayMerge}, \code{heapSort}, \code{a-star}, and \code{prim}, \proj-O2 incurs performance degradation, but not in \proj. 
For example, \code{prim} shows 8\% performance gain under \proj. However, it incurs 33\% performance loss when we turn on backend optimizations in \proj-O2. 
\YL{\code{prim}'s compute kernel has a control flow that checks if three conditions are all met.}
\YL{Unwantedly, the downstream optimizer sees our MERIT-transformed code as "unoptimized" \code{select} and reverts back to branches, one for each condition check. Thus, compared to the original branch-ed code that has only one jump target, \proj-O2 backend generates three jump targets, unnecessarily leading to more hardware mispredictions.}

\noindent\fbox{\parbox{0.99\columnwidth}{
\textit{\textbf{Takeaway}}: Since \proj operates on early LLVM-IR level, existing downstream passes might transform the code suboptimally, including reverting branchless to branched code.
}}

\textit{\textbf{Early if-conversion}} shows much less performance improvement (only 1.06x geomean).
This is because it cannot transform control-flow regions where unsafe memory operations exist.
For example, \code{toUpper} has \code{str[i]} within the branch (Listing~\ref{lst:to_upper}) that might cause invalid memory access on x86. However, such a problem is avoided in \proj by safely guarding memory access during IR instruction melding. 

We noticed a rare case in \code{dutchFlag} where both \proj and if-conversion show improvement, but if-conversion beats \proj. 
This algorithm's loop contains a highly unpredictable, three-way branch that guards only simple counter increments. Early if-conversion excels here by transforming this into computationally trivial \code{cmov} instructions, eliminating the misprediction penalties.
This specific workload is ideal for if-conversion because it lacks the complex, unsafe memory operations that \proj is designed to handle. 
Furthermore, as a later-stage pass, the extremely low register pressure and the simple data dependency chains mean that backend optimizations (register allocation, instruction scheduling) work well with early if-conversion's output, whereas \proj's IR-level \code{selects} may undergo additional lowering transformations that introduce relative inefficiencies.

\begin{figure}[t]
    \centering
    \begin{subfigure}{\columnwidth}
        \centering
        \includegraphics[width=0.99\columnwidth]{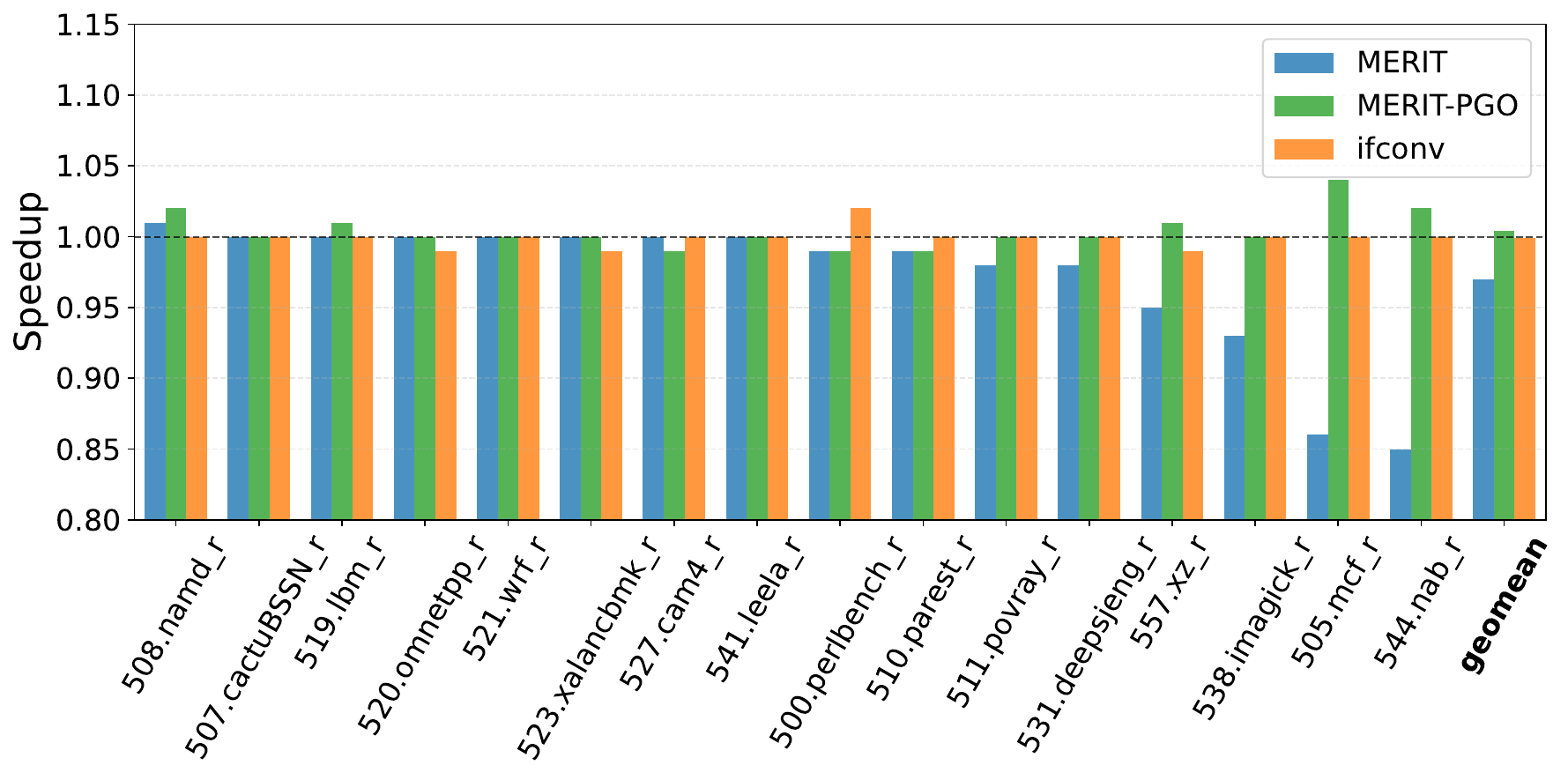}
        \caption{\proj achieves performance geomean of 0.97, \proj-PGO of 1.01, compared to if-conversion's 1.}
        \label{fig:rate_perf}
    \end{subfigure}
    \vspace{1em}
    \begin{subfigure}{\columnwidth}
        \centering
        \includegraphics[width=0.99\columnwidth]{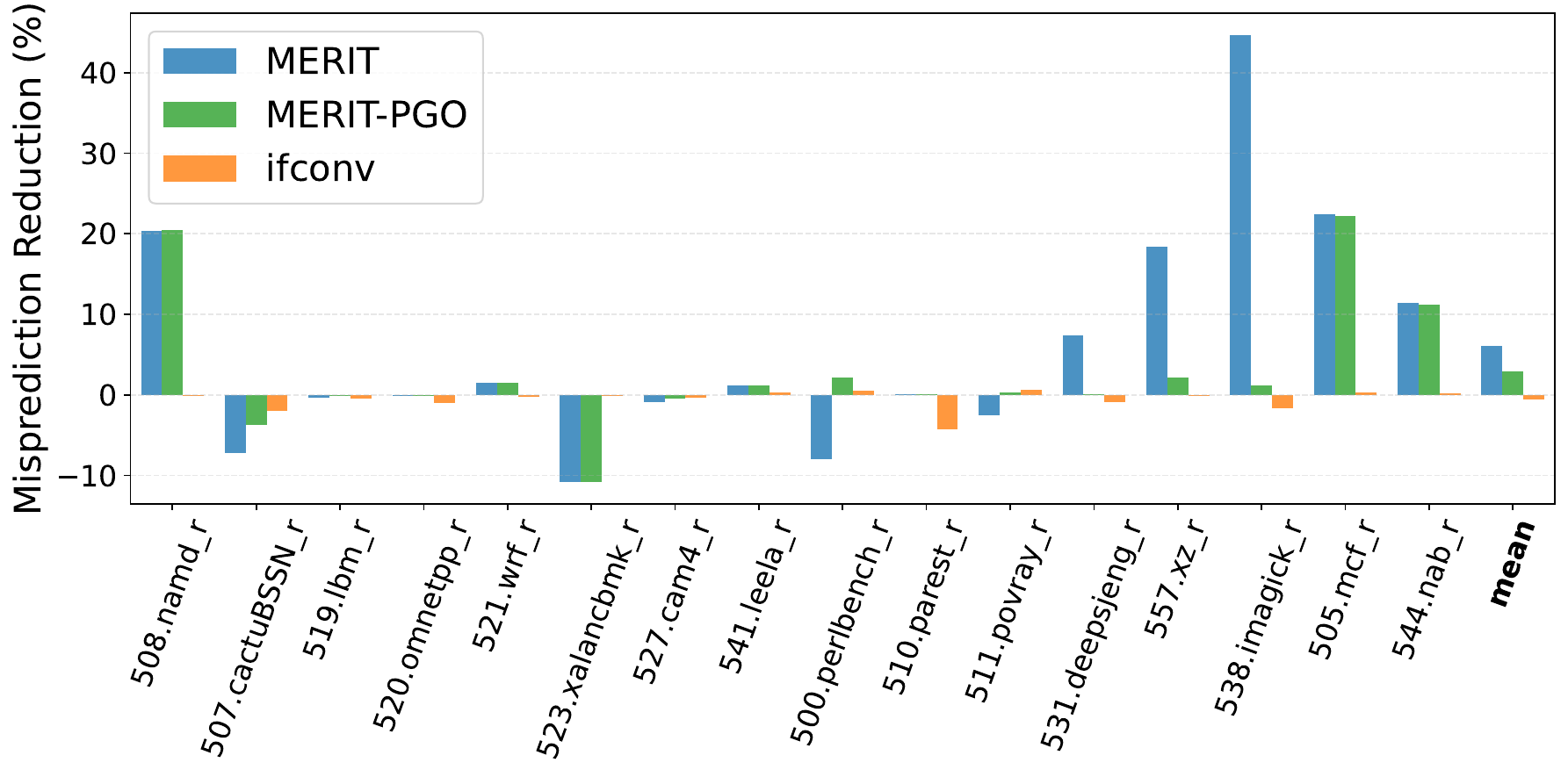}
        \caption{\proj achieves average 6.1\%, \proj-PGO achieves 3\% compared to if-conversion's -0.6\% branch misses reduction.}
        \label{fig:rate_br_miss_reduction}
    \end{subfigure}
    \vspace{-2em}
    \caption{Performance and branch misses reduction of \proj and early if-conversion on \textbf{SPECrate 2017}.}
    \label{fig:rate_perf_br_miss}
\end{figure}

\paragraph{CPU SPEC2017 Performance}
Figure~\ref{fig:rate_perf} shows SPECrate 2017's performance transformed by \proj's pass.
Unfortunately, blindly applying \proj on most benchmarks shows performance degradation (blue bars). This is because most benchmarks are not branch-heavy, thus \proj produces excessive runtime instructions, giving geomean of 0.97 compared to no transformation.
However, when we apply PGO to identify and exclude those functions that cause most performance degradation, all benchmarks show either the same or better performance (green bars). 
Take \code{505.mcf\_r} as an example, naive \proj generates 14\% degradation but shows 4\% runtime improvement when we identify and exclude those bad functions.
If-conversion, the same behavior we've seen in microbenchmarks, barely shows any performance differences, indicating its conservativeness when producing branchless code.


\paragraph{Misprediction Reduction}
We evaluate how well \proj eliminates branch mispredictions compared to if-conversion. 
On average, \proj reduces 48\% branches mispredictions on the microbenchmarks (Figure~\ref{fig:dse_br_miss_reduction}) and 6.1\% on SPECrate 2017 (Figure~\ref{fig:rate_br_miss_reduction}), respectively. 
Compared to early if-conversion, \proj reduces 40.8\% more mispredictions in our microbenchmarks and 6.7\% more in SPECrate 2017. 
For the microbenchmarks, the overall trend of misprediction reduction matches the performance improvement in Figure~\ref{fig:dse_perf}.
However, for a more realistic SPECrate benchmark, misprediction reductions do not directly correlate to performance improvement. 
For example, in \code{505.mcf\_r}, although \proj and \proj-PGO reduce same amount of mispredictions, \proj-PGO gives 4\% positive performance compared to \proj who gives 14\% performance degradation. 
In \code{538.imagick\_r}, although \proj reduces 44\% mispredictions, it counterintuitively drags down the performance by 7\%.

\noindent\fbox{\parbox{0.99\columnwidth}{
\textit{\textbf{Takeaway}}: For complex workloads, solely focusing on reducing misprediction does not necessarily result in better performance. It is beneficial to have an intelligent cost model to avoid applying optimizations that lead to heavy overhead.
}}



\begin{figure}[t]
    \centering
    \begin{subfigure}{\columnwidth}
        \centering
        \includegraphics[width=0.99\columnwidth]{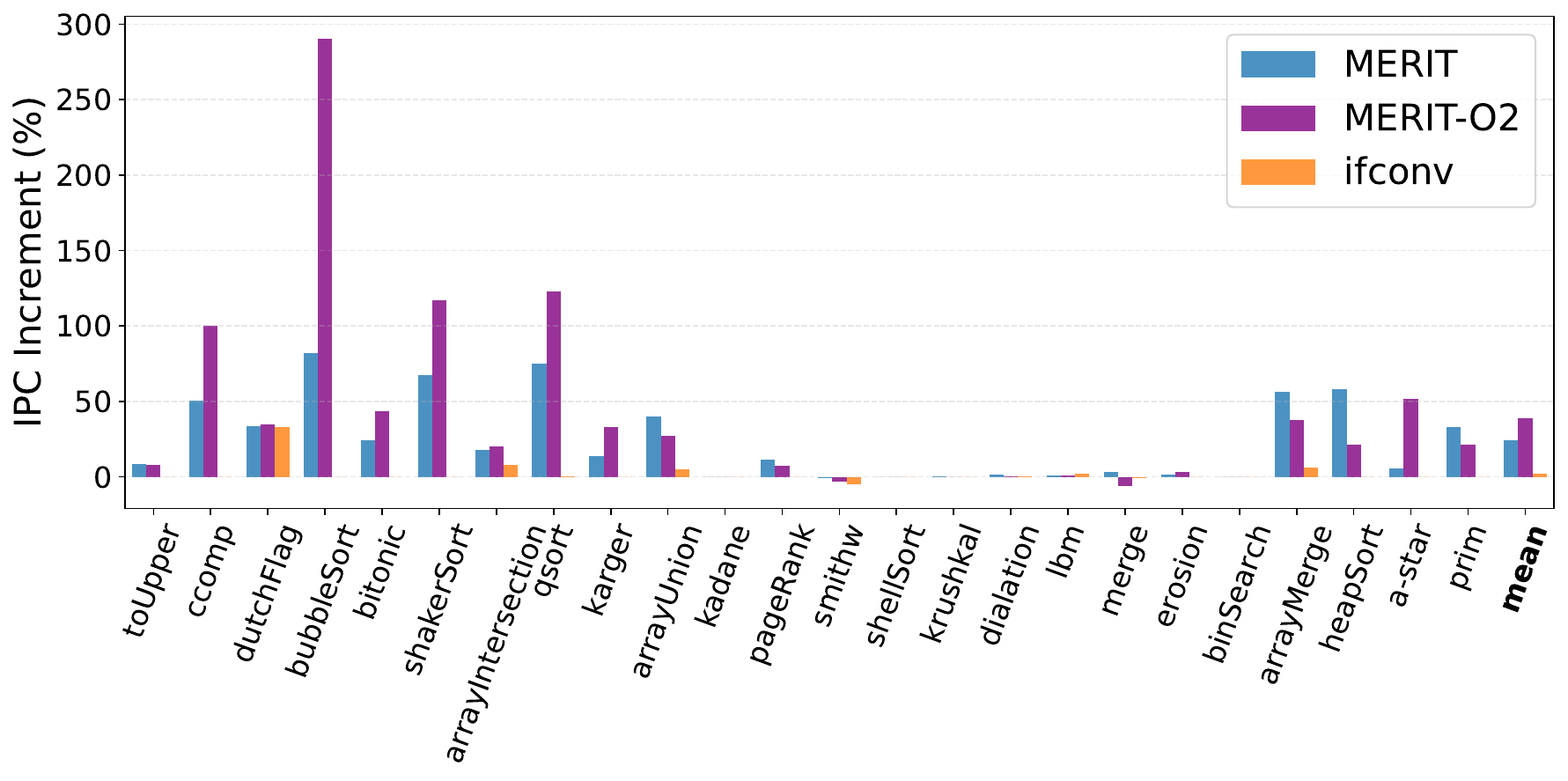}
        \caption{\proj achieves 24.4\% IPC improvement compared to if-conversion's 2.1\% on average.}
        \label{fig:dse_ipc_inc}
    \end{subfigure}
    \vspace{1em}
    \begin{subfigure}{\columnwidth}
        \centering
        \includegraphics[width=0.99\columnwidth]{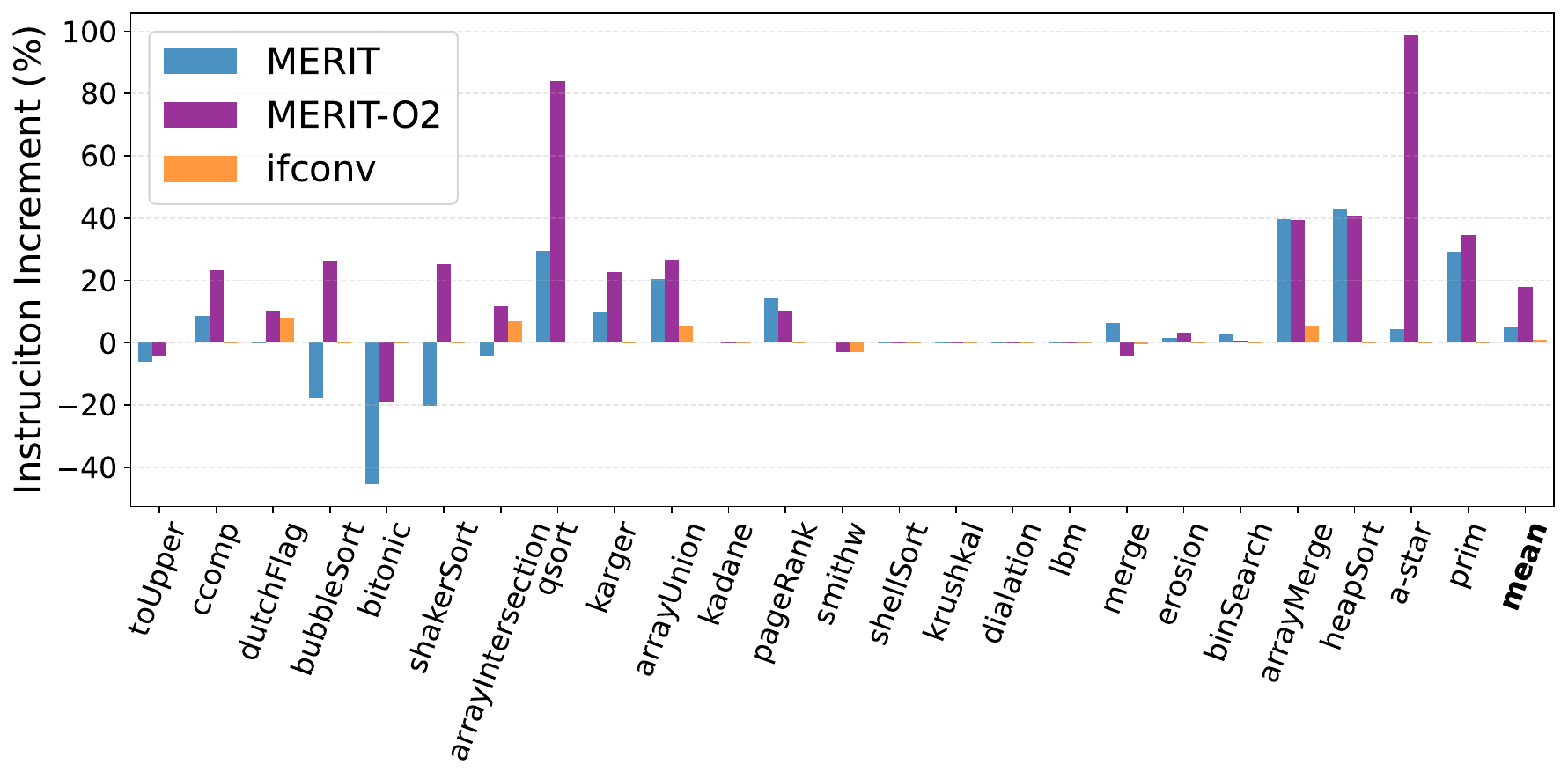}
        \caption{\proj incurs 17.8\% runtime instruction overhead compared to if-conversion's 0.9\% on average.}
        \label{fig:dse_instruction_inc}
    \end{subfigure}
    \vspace{-2em}
    \caption{IPC increment and runtime instruction overhead of \proj and early if-conversion on \textbf{microbenchmarks} compared to no transformation.}
    \label{fig:dse_ipc_io}
\end{figure}

\begin{figure}[t]
    \centering
    \begin{subfigure}{\columnwidth}
        \centering
        \includegraphics[width=0.99\columnwidth]{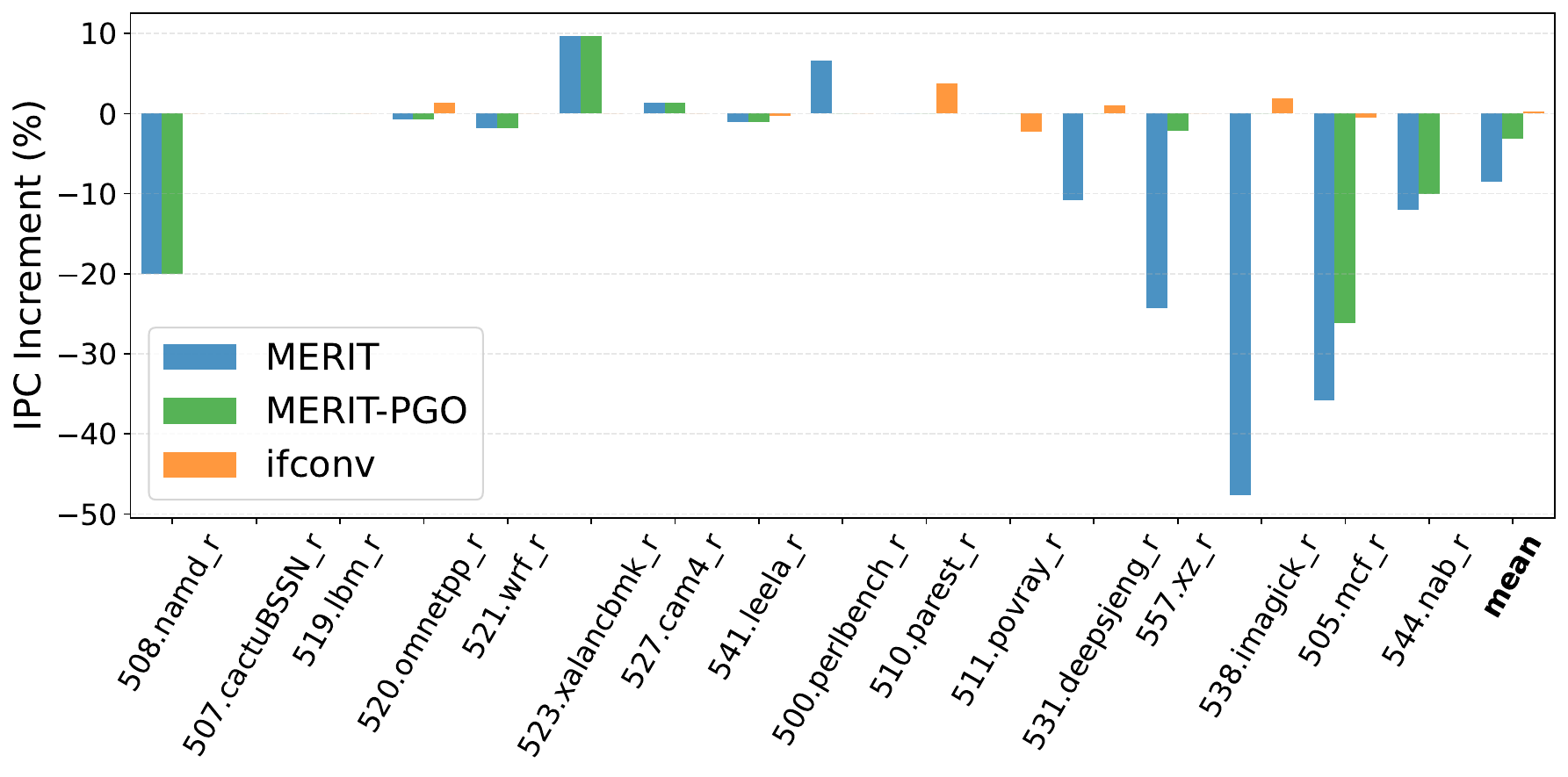}
        \caption{\proj incurs 8.5\% IPC degradation. \proj-PGO incurs 3.2\% IPC degradation, compared to if-conversion's 0.3\% IPC improvement.}
        \label{fig:rate_ipc_inc}
    \end{subfigure}
    \vspace{1em}
    \begin{subfigure}{\columnwidth}
        \centering
        \includegraphics[width=0.99\columnwidth]{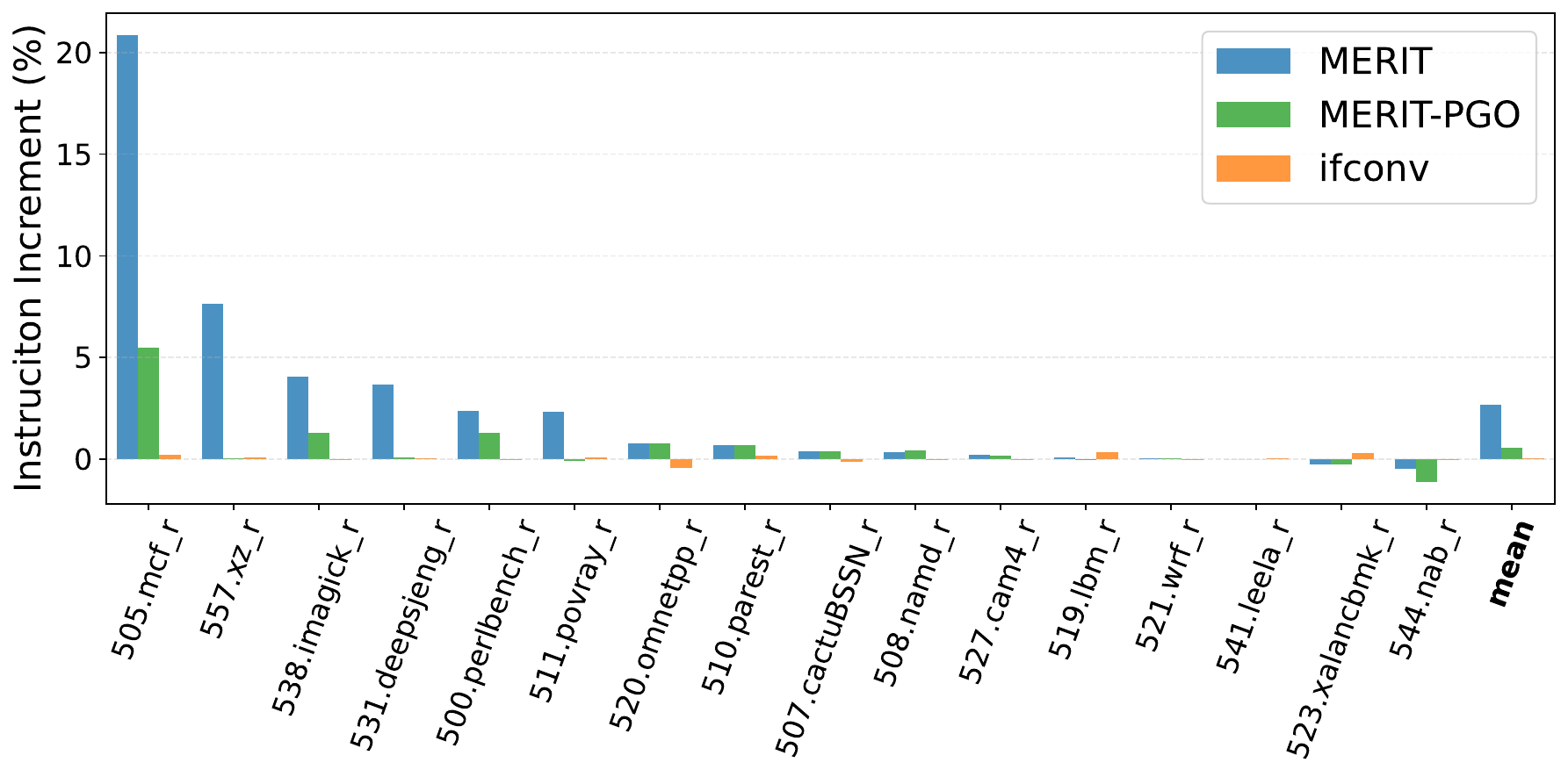}
        \caption{\proj incurs 2.7\% runtime instruction overhead, \proj-PGO incurs 0.6\% overhead compared to if-conversion's 0.1\% overhead on average.}
        \label{fig:rate_instruction_inc}
    \end{subfigure}
    \vspace{-2em}
    \caption{IPC increment and runtime instruction overhead of \proj and early if-conversion on \textbf{SPECrate 2017}.}
    \label{fig:rate_ipc_io}
\end{figure}

\paragraph{IPC improvement}
Instruction per cycle (IPC) directly reflects how compilers transforms programs to take advantage of processor's ILP.
As shown in Figure~\ref{fig:dse_ipc_inc}, \proj and \proj-O2 significantly improve IPC for the microbenchmarks.
However, for the complex SPECrate workloads (Figure~\ref{fig:rate_ipc_inc}), the relationship between IPC and performance is more nuanced. For example, in \code{508.namd\_r} and \code{505.mcf\_r}, \proj-PGO achieves speedups of 2\% and 4\%, respectively (Figure~\ref{fig:rate_perf}) , despite showing a decrease in IPC of 20\% and 26\% (Figure~\ref{fig:rate_ipc_inc}).

This apparent paradox is explained by the total number of instructions being retired. The baseline (no transformation) suffers from frequent mispredictions, forcing the CPU to speculatively execute and retire a massive number of instructions down the wrong path. \proj eliminates these mispredicted paths entirely. While \proj adds a small number of extraneous instructions to the correct path, this increase is far smaller than the number of wrong-path instructions being eliminated. Therefore, the total number of dynamic instructions retired (Instructions\_new) decreases significantly. Both total cycles (Cycles\_new) and total instructions (Instructions\_new) are reduced, but instructions are reduced more than cycles, resulting in a lower IPC (Instructions\_new / Cycles\_new) even as the total execution time drops (a speedup).

\noindent\fbox{\parbox{0.99\columnwidth}{
\textit{\textbf{Takeaway}}:
\proj's insertion of extraneous instructions is only one part of the story. Its primary benefit comes from eliminating the much larger instruction overhead of mispredicted paths. In complex workloads, this can lead to a net decrease in total dynamic instructions, causing IPC to drop even as wall-clock time improves.
}}

\paragraph{Dynamic Instruction Overhead}
The key consideration for branch elimination is the growth in dynamic instructions. Our transformation, by design, converts control flow into data flow, which often causes instructions from both paths of the original branch to be executed. Figures~\ref{fig:dse_instruction_inc} and~\ref{fig:rate_instruction_inc} quantify this overhead for the microbenchmarks and SPECrate suites, respectively.

On our microbenchmarks, many workloads show a 10\% to 100\% increase in dynamic instructions. The \code{a-star} benchmark is a clear example where this overhead contributes to its slowdown. 
However, this overhead does not always correlate with a performance loss.
Workloads like \code{qsort}, \code{arrayMerge}, and \code{heapSort} all see performance gain despite the runtime instruction overhead. This means that the cost of these extra instructions is negligible compared to the penalty of frequent, severe branch mispredictions.

On the SPECrate suite (Figure~\ref{fig:rate_instruction_inc}), naive \proj incurs an average overhead of 2.7\%. This highlights the importance of \proj-PGO, which reduces this overhead to just 0.6\% by selectively disabling the transformation on functions where the instruction overhead would not be offset by misprediction-elimination gains.

\subsection{Case Study: SQLite}
Our first case study, a \proj-compiled SQLite engine running 22 TPC-H queries, clearly demonstrates the necessity of profile-guided selective transformation.

\begin{figure}[t]
    \centering
    \includegraphics[width=0.99\columnwidth]{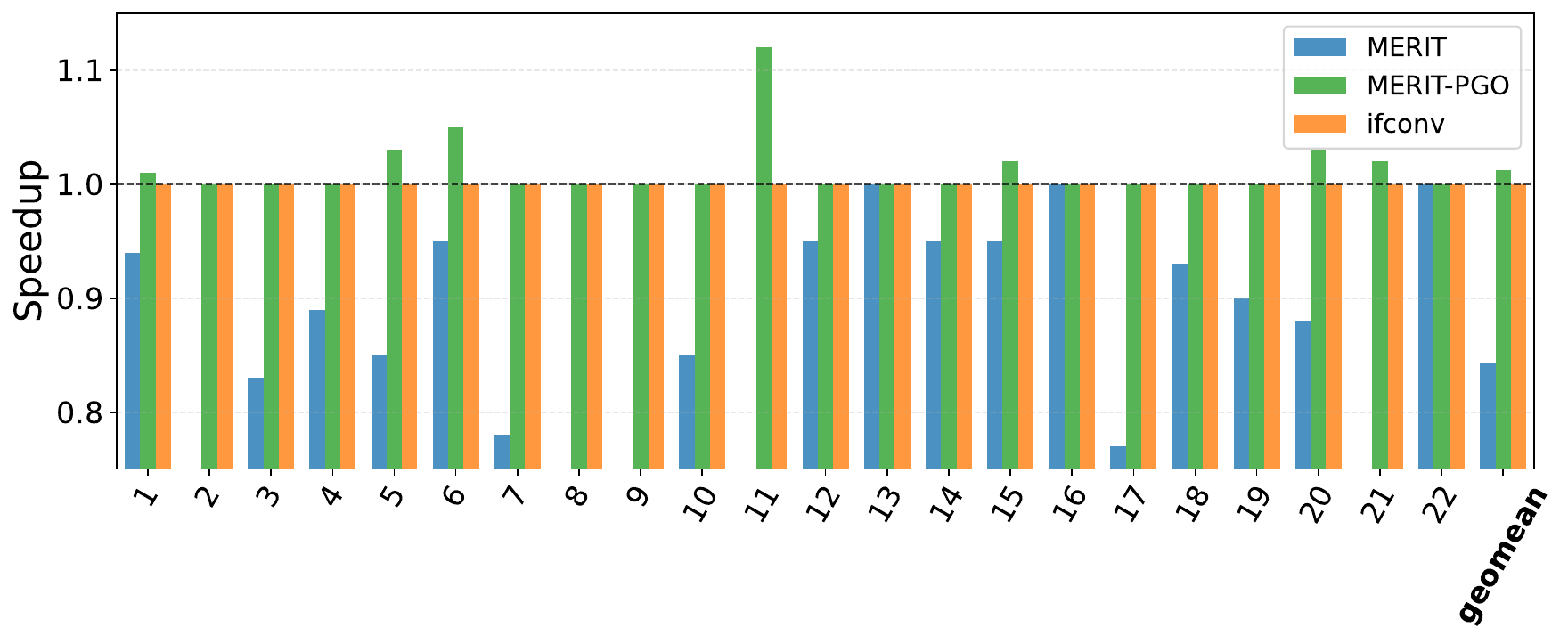}
    \caption{\proj achieves performance geomean of 0.84, \proj-PGO of 1.01, compared to if-conversion's 1.}
    \label{fig:sqlite_perf}
\end{figure}

As shown in Figure~\ref{fig:sqlite_perf}, naively applying \proj (blue bars) results in significant performance degradation on many queries, particularly 7, 8, 9, and 21, leading to a geomean of 0.84. This is again due to the instruction overhead from transforming branches that were not performance bottlenecks.

However, the \proj-PGO (green bars) results are transformative. By using PGO to identify and exclude problematic functions, we completely mitigate all slowdowns and achieve a positive geomean performance of 1.01. Furthermore, \proj-PGO achieves speedups on several queries (e.g., 5, 6, 11, 15, and 18) by targeting only the high-value, misprediction-prone branches. This selective transformation reinforces our takeaway: for complex real-world applications, \proj must be paired with an intelligent cost model or PGO filtering to achieve performance gains.
If-conversion does not make any performance differences, likely because it cannot transform performance-critical blocks.

\subsection{Case Study: Python}
\begin{figure}[t]
    \centering
    \includegraphics[width=0.99\columnwidth]{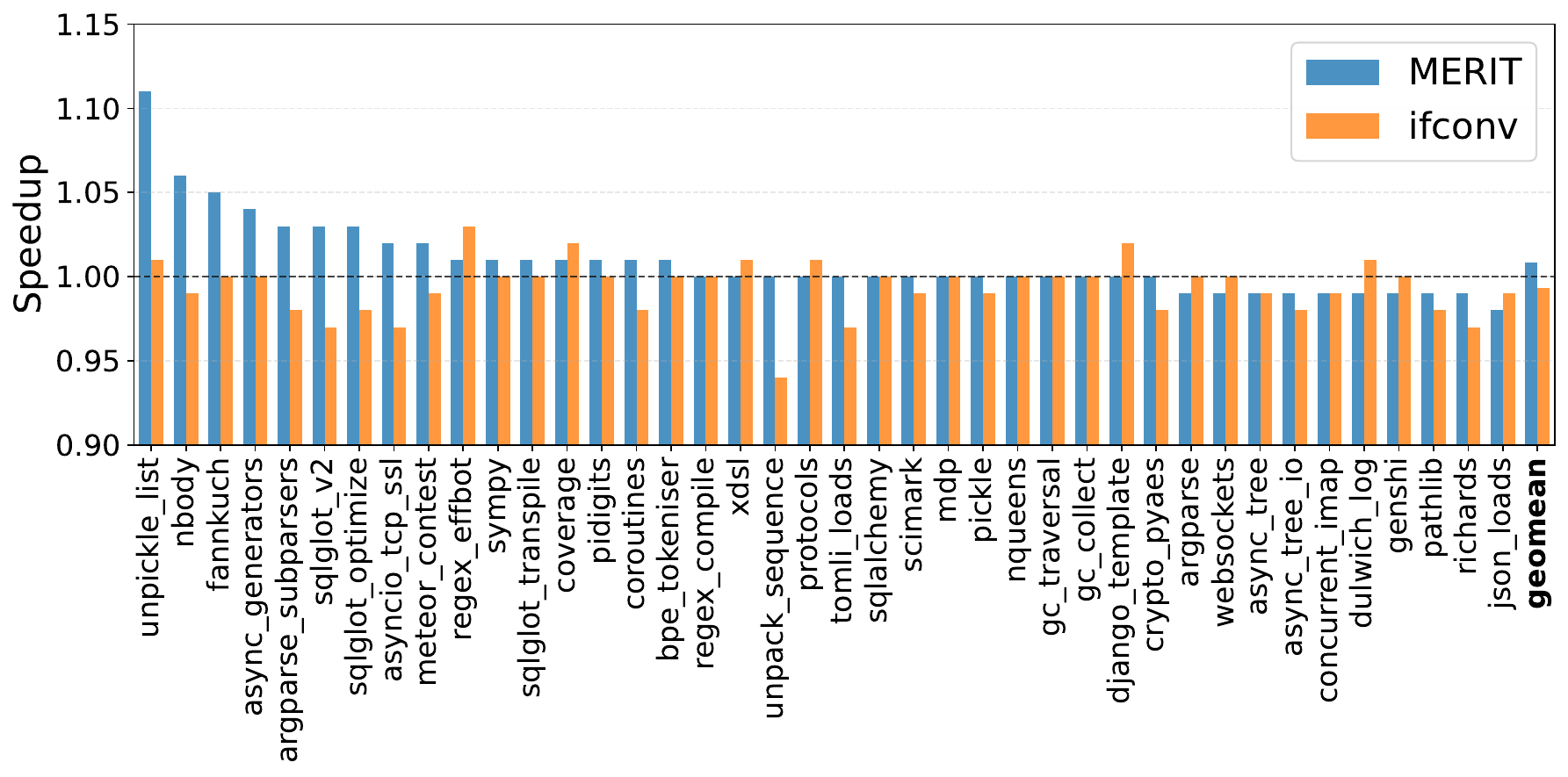}
    \caption{\proj achieves 1.01 geomean performance compared to if-conversion's 0.99.}
    \label{fig:pyperformance_perf}
\end{figure}

In our second case study, we evaluate \proj on a large real-world application: the CPython interpreter, with 40 benchmarks from the pyperformance suite~\cite{pyperformance}. Shown in Figure~\ref{fig:pyperformance_perf}, applying \proj indiscriminately results in a 1.01 geomean speedup, with some workloads like \code{unpickle\_list} and \code{nbody} seeing benefits, while others do not. In comparison, the standard if-conv pass is more volatile, showing gains on some workloads (e.g., \code{regex\_effbot}) but significant losses on others (\code{unpack\_sequence}), for a 0.99 geomean.

We do not report \proj-PGO performance for the Python suite because its function-level filtering is too coarse-grained for a complex runtime like the interpreter, where diverse workloads from the pyperformance suite stress different parts of the code. Excluding a function to benefit one workload (\eg \code{nbody}) could harm another (\eg \code{json\_loads}) that relies on the same function's transformation. However, this points to a practical strategy for specialized environments: one could use \proj with PGO to create a highly-tuned interpreter optimized only for their specific workload (e.g., a JSON serialization service), transforming only the hot, misprediction-prone functions in their critical path to achieve speedups that would be washed out in a general-purpose benchmark.



%% file: sections/discussion.tex
\section{Limitation and Future Work}
\label{sec:discussion}

We identified two key limitations that constrain \proj's effectiveness: unavailability of a static cost model for selective transformation and a lack of coordination between \proj and backend optimization passes.

\paragraph{Need for Cost Model}
\YL{The current scope of \proj is limited: it applies the transformation uniformly to any eligible branch if the IR alignment score meets the threshold.} Such coarse application can be counterproductive, especially when a branch is already highly predictable (e.g., loop-invariant) or when merging control-flow paths introduces excessive instruction overhead.
Developing a fine-grained cost model to estimate the profitability is a non-trivial challenge because \proj operates at the target-independent IR level, lacking the scheduling and latency information available to late-stage Machine-IR passes. 

The future work is \YL{to come up with a heuristic-based \proj transformation cost model, to uniformly consider the IR alignment score, the resulting static instruction overhead, and the data-flow dependency depth of the new \code{select} chain.
This might require aggressively applying \proj in early LLVM-IR and partially reversal to branched code in late machine IR, depending on the resource subscription only available in late compile stage~\cite{partial_reverse}.}

However, one fundamental challenge still exists: the compiler must know whether the hardware's branch predictor will fail on certain hard-to-predict branches. 
\YL{This requires fine-grained PGO than the current naive one, to identify per-branch misprediction penalty.}
\YL{If PGO is not applicable, another promising direction is to explore using Large Language Models (LLMs), or transformer-based execution estimation models like TRACED~\cite{traced} to create a powerful, AI-driven cost model, which can be fine-tuned on vast codebases to recognize complex patterns that simple heuristics might miss. }
Such an approach aims to move \proj closer to zero-cost, highly effective static optimization.

\paragraph{Lack of \proj-aware Backend Optimizations}
The second limitation is the lack of coordination and conflicting goals between \proj and backend optimization passes.  
Downstream passes may {undo} \proj's work or make suboptimal decisions.
We observed this as a "performance anomaly" where aggressive backend optimizations sometimes provided a greater relative benefit to the original branching code, masking \proj's true potential.

This can happen for several reasons: later passes may be expecting code with branches such that they can apply certain optimizations like register pressure reduction or target-specific predication. This could lead to undoing of \proj's transformation.
Future work will address this by using LLVM metadata to pass compiler hints from \proj to backend passes to preserve the branchless structure.


%% file: sections/relatedwork.tex
\section{Related Work}
\label{sec:related}

Several techniques have been proposed to address the branch prediction including leveraging Fetch Directed Instruction Prefetching (FDIP)~\cite{fdip, rebase_ins_prefetch}, 
TAGE-like~\cite{tage, tage_64, tage_sc_l, tage_sc_l_again}, 
or perceptron-based~\cite{perceptron_0, perceptron_1}.
Despite their aim for high accuracy, they struggle with noisy histories and are often overwhelmed by large branch footprints of data center applications, leading to frequent capacity-induced misses.

Hybrid approaches leverage software PGO, a technique widely used in data centers where application profiles are collected in production and used for offline analysis~\cite{lightning_bolt, bolt, autofdo, asmdb, twig}. 
Whisper~\cite{whisper} analyze offline profiles to generate lightweight Boolean formulas that are injected into the binary as prediction hints for the hardware.
BranchNet~\cite{zangeneh2020branchnet} uses offline profiling to train complex machine learning models, which are then fed to runtime via a small on-chip inference engine to predict most difficult branches. 
These techniques augment the hardware predictor by using offline compute resources to solve prediction problems that are intractable for hardware to learn at runtime.

%% file: sections/conclusion.tex
\section{Conclusion}
\label{sec:conclusion}


This paper introduces \proj, a novel compiler transformation that eliminate branches on x86 by melding structurally similar operation sequences.
Unlike traditional if-conversion, \proj's "select operands, compute once" strategy uses semantic analysis for safe operand-level guarding, enabling it to safely transform branches with conditional memory operations that other methods must skip. 
We also demonstrate how \proj can be easily integrated to PGO by allowing selective transformation. \proj achieves on average 10.9\% speedups (up to $32\times$) over 102 workloads. 
Our analysis reveals that \proj's effectiveness is currently constrained by the lack of a static cost model and by "\proj-blind" backend optimizations that can undo its work, highlighting these as critical areas for future research.